\documentclass[conference]{IEEEtran} 
\IEEEoverridecommandlockouts

\usepackage{cite}
\usepackage{amsmath,amssymb,amsfonts}
\usepackage{algorithmic}
\usepackage{graphicx}
\usepackage{textcomp}
\usepackage{xcolor}
\usepackage{hyperref}
\usepackage{subcaption}
\usepackage{adjustbox}
\usepackage{multirow}

\def\BibTeX{{\rm B\kern-.05em{\sc i\kern-.025em b}\kern-.08em
    T\kern-.1667em\lower.7ex\hbox{E}\kern-.125emX}}
\begin{document}

\title{Selection of Layers from Self-supervised Learning Models for Predicting Mean-Opinion-Score of Speech\\
\thanks{The computations handling was enabled by the supercomputing resource Berzelius provided by National Supercomputer Centre at Link\"oping University and the Knut and Alice Wallenberg foundation.}
}









\author{
    \IEEEauthorblockN{Xinyu Liang \IEEEauthorrefmark{1}, Fredrik Cumlin \IEEEauthorrefmark{2}, Victor Ungureanu
    \IEEEauthorrefmark{3}, Chandan K. A. Reddy \IEEEauthorrefmark{3}, Christian Sch\"uldt \IEEEauthorrefmark{3}, Saikat Chatterjee \IEEEauthorrefmark{2}} 

    \IEEEauthorblockA{\IEEEauthorrefmark{1} HCLTech AB, Sweden, \,\,\, \IEEEauthorrefmark{3} Google LLC}
    \IEEEauthorblockA{\IEEEauthorrefmark{2} School of Electrical Engineering and Computer Science, KTH Royal Institute of Technology, Sweden}
    \IEEEauthorblockA{\IEEEauthorrefmark{1} hopeliang@icloud.com,
    \IEEEauthorrefmark{2} \{fcumlin, sach\}@kth.se, \IEEEauthorrefmark{3} \{ungureanu, chandanka, cschuldt\}@google.com}
}

\maketitle

\begin{abstract}
Self-supervised learning (SSL) models like Wav2Vec2, HuBERT, and WavLM have been widely used in speech processing. These transformer-based models consist of multiple layers, each capturing different levels of representation. While prior studies explored their layer-wise representations for efficiency and performance, speech quality assessment (SQA) models predominantly rely on last-layer features, leaving intermediate layers underexamined. In this work, we systematically evaluate different layers of multiple SSL models for predicting mean-opinion-score (MOS). Features from each layer are fed into a lightweight regression network to assess effectiveness. Our experiments consistently show early-layers features outperform or match those from the last layer, leading to significant improvements over conventional approaches and state-of-the-art MOS prediction models. These findings highlight the advantages of early-layer selection, offering enhanced performance and reduced system complexity.

\end{abstract}


\section{Introduction}

Self-supervised learning (SSL) models, trained on large-scale datasets with numerous parameters, have demonstrated considerable generalization abilities across various applications, including image, speech, audio, and natural language processing. In the vast field of speech processing, state-of-the-art SSL models like Wav2Vec2 \cite{wav2vec2}, HuBERT \cite{hubert}, and WavLM \cite{wavlm} have emerged as powerful feature extractors. Features extracted from these SSL models have been applied to many types of downstream tasks, such as automatic speech recognition (ASR) \cite{ssl_asr}, audio classification \cite{ssl_ac}, speaker verification (SV) \cite{ssl_sv}, etc. The mentioned SSL models are predominantly transformer architectures-based, having many layers in deep learning systems. Conceptually, all the layers can provide features with varied and complementary information.

Recent studies have investigated the general layer-wise behavior of speech SSL models. Chung et al. quantitatively measured the similarity between the last-layer representations for different SSL models, and found that pre-training loss plays a critical role in downstream tasks \cite{chung2021similarity}. Pasad et al. examined different layers of a Wav2Vec2 model on acoustic and linguistic content by performing canonical correlation analysis (CCA) and measuring an information-theoretic quantity called mutual information (MI) \cite{pasad2021layer}. Later, they expanded the study for various SSL models \cite{pasad2023comparative}. One of the key findings in their work is the consistent trend observed across all popular speech SSL models: as the latent features progress through the transformer layers, the CCA similarity with spectrogram features decreases, while the similarity with phoneme and word labels increases.

Sajjad et al. \cite{sajjad2020poor} studied different layer dropping methods of stacked transformer models and found that top-layer dropping can both improve efficiency and performance. In the speech domain, analyses have shown that layer selection is essential depending on the downstream task when adopting pre-trained SSL models \cite{fan2020exploring, wavlm, SUPERB}. However, for speech quality assessment (SQA), a much-exercised approach to apply SSL models' generated features is to use the features from the last layer, and then feed the features to a shallow network. 

\noindent\textbf{Our contributions:} In this paper we study the suitability of different layers from the pre-trained speech SSL models for predicting mean-opinion-score (MOS) of speech. We evaluate six speech SSL models of varying sizes and training methodologies, conducting experiments on three benchmark datasets that encompass three languages and diverse scenarios. Our analysis examines individual layers by feeding their extracted features into a projection head - a lightweight inference network, designed to accommodate limited MOS-labeled training data, together with a $\ell_2$-norm-based loss function. We observe a consistent trend shared across all scenarios: features extracted from early layers of the SSL models provide better MOS prediction performance than the features extracted from the last layer. Our findings inform the selection of optimal early layers. The proposed system, leveraging early-layer features and a simplified design, outperforms state-of-the-art MOS prediction models that rely on last-layer features with complicated system design and limited data applicability. Additionally, we explore fine-tuning of the pre-trained SSL models, and experimental results corroborate with the same hypothesis.

\subsection{Relevant literature on MOS prediction}

MOS from subjective listeners is the gold standard for comprehensive SQA, though it is costly and time-consuming. Advances in deep neural network (DNN) techniques have enabled the development of non-intrusive methods to predict human MOS. Contemporary work in this field follows two main paradigms: end-to-end trained models and SSL-based approaches.

End-to-end models, which are small enough for real-time SQA tasks, have been explored with varying complexities and objective functions \cite{DNSMOS, DNSMOSp, MOSNet, DeePMOS, DeePMOSb, LDNet, NISQA}. However, these models, limited by the size of their training data and the number of parameters, often face challenges with generalization, for example when tested on out-of-domain data \cite{SSL-MOS}.

State-of-the-art (SOTA) models leverage features extracted from SSL models to enhance their performance. SSL-MOS \cite{SSL-MOS} was the first to demonstrate the benefits of using SSL features for MOS prediction. Subsequent models, such as ZevoMOS, UTMOS, and LE-SSL-MOS \cite{ZevoMOS, UTMOS, LE-SSL-MOS}, sought to extract more information from the datasets, incorporating individual ratings, rater identities, and additional data from other systems like ASR. All these models primarily utilize the final layer output from one or multiple SSL models. To date, there exists no experimental study that has explored the potential benefits of using different layers of a chosen SSL model to determine whether the layers contain relevant information for better MOS prediction performance. We perform such a study in this paper.

\section{Selection of Layers in SSL models for MOS Prediction}
\label{sec:method}

\begin{figure}
    \centering
    \includegraphics[width=1\linewidth]{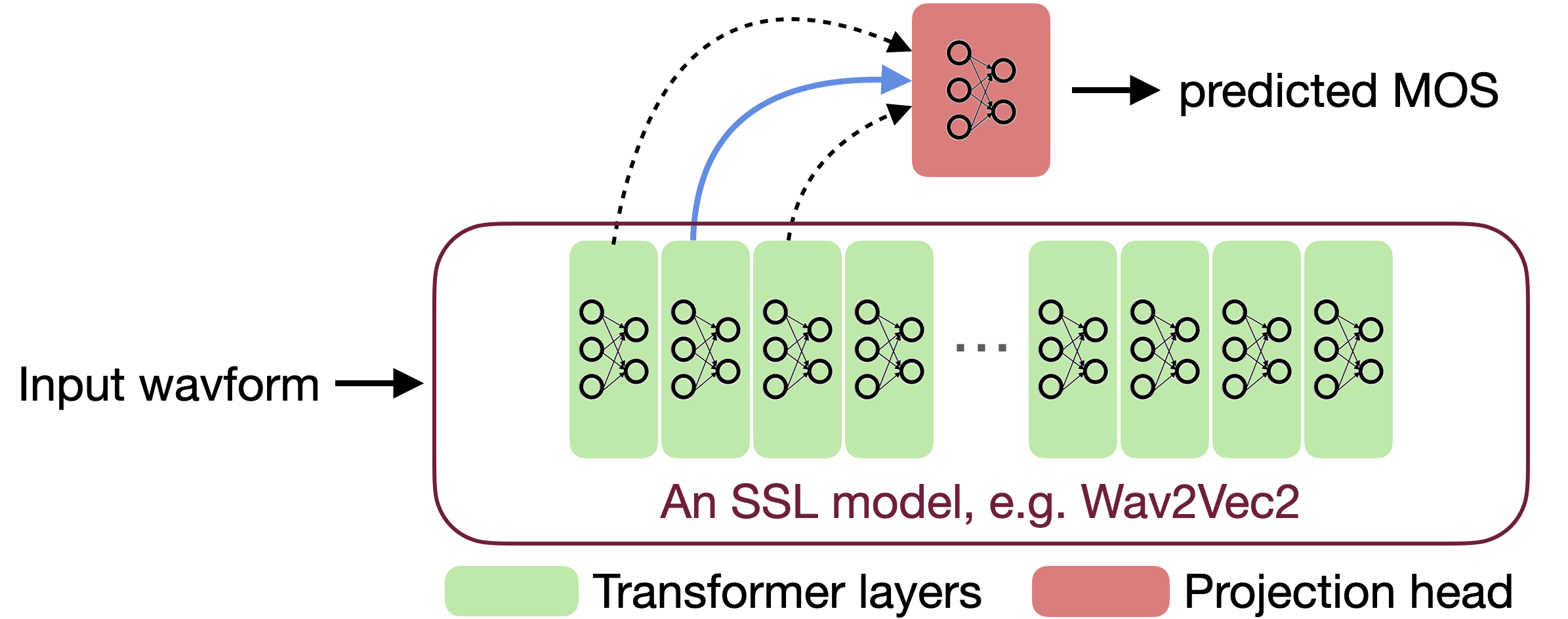}
    \caption{System diagram for evaluating layers of an SSL model. Here the `solid thick arrow', connecting the SSL model and the projection head, represents signal flow (or features use) from a chosen layer of the SSL model to the projection head. A `dotted arrow' represents the idea that it can be a `solid arrow' if the corresponding layer is chosen. Note that we choose one layer at a time in our experiment.}
    \label{fig:system-diagram}
\end{figure}

We evaluate several SSL models. To do this, we design an exhaustive search-based system to evaluate the suitability of different layers from an SSL model for the task of MOS prediction, analyzing each layer one by one. This is achieved by attaching a projection head model to each transformer layer's output, and train it using MOS-labeled datasets. The overall system is illustrated in Figure \ref{fig:system-diagram}. Each transformer layer's output is of shape [$T$, $D$], where $T$ denotes the number of time frames, and $D$ is the feature dimension. The architecture of the projection head is motivated by the DNSMOS architecture \cite{DNSMOS} and was adopted similarly as in 
\cite{MultiGauss}. A major argument to evaluate the layers one by one, instead of an automatic process or fusing them together, is motivated by the much-cited study of layer-wise investigation of convolutional neural networks for image processing and recognition \cite{zeiler2014visualizing}.

\subsection{Speech SSL models}

To conduct our experiments, we utilize several widely adopted pre-trained speech SSL models, including Wav2Vec2, HuBERT, and WavLM~\cite{wav2vec2, hubert, wavlm}. Wav2Vec2 extracts features using a convolutional encoder, followed by transformer layers for context modeling. During pre-training, it masks encoded speech/audio segments and predicts them, allowing the model to learn useful representations without labeled data. HuBERT builds on Wav2Vec2 by incorporating a clustering step during pre-training. Instead of predicting masked raw speech features, HuBERT clusters hidden audio representations and then predicts these clusters for the masked regions. WavLM further extends Wav2Vec2 and HuBERT by introducing a gated relative position bias in the transformer blocks and employing a multi-task learning method during the pre-training, by adding a denoising objective along with self-supervised prediction. 


In our study, we consider several pre-trained SSL models for general-purpose tasks. The specifics of the selected SSL models \footnote{https://pytorch.org/audio/stable/pipelines.html} are summarized in Table \ref{tab:ssl_models}.

\begin{table}[h!]
\small
\centering
\begin{tabular}{l|cccc}
\hline
\textbf{model name}        & \textbf{$\#$layers} & \textbf{$\#$dim} & \textbf{$\#$params} & \textbf{train data}          \\ \hline
w2v2\_base          & 12   & 768  & 94.4M   & 960hr      \\
hubert\_base        & 12   & 768  & 94.4M   & 960hr      \\ 
wavlm\_base         & 12   & 768  & 94.4M   & 960hr      \\
w2v2\_xlsr\_300m    & 24   & 1024 & 315M    & 436,000hr  \\
w2v2\_xlsr\_1b      & 48   & 1280 & 962M    & 436,000hr  \\
w2v2\_xlsr\_2b      & 48   & 1920 & 2.16B   & 436,000hr  \\ \hline
\end{tabular}
\caption{Specifications of selected SSL models}
\label{tab:ssl_models}
\end{table}

\subsection{Datasets for MOS prediction}

For MOS prediction, commonly referred benchmark datasets are VoiceMOS Challenge 2022 (BVCC) \cite{BVCC} dataset, the Tencent corpus \cite{Tencent}, and the NISQA corpus \cite{NISQA}. These three datasets encompass different languages and a variety of audio scenarios such as noisy environments. We summarize the key details of the three datasets in Table \ref{tab:datasets}.

\begin{table}[h!]
\centering
\begin{adjustbox}{width=\columnwidth}
\begin{tabular}{l|ccccc}
\hline
\multirow{2}{*}{\textbf{Dataset}} & \multirow{2}{*}{\textbf{language}} & \multirow{2}{*}{\textbf{Raters$/$Clip}} & \multicolumn{3}{c}{\textbf{Number of clips}} \\
& & & Train & Validation & Test \\ \hline
BVCC & English & 8 & 4,974 & 1,066 & 1,066 \\
Tencent & Chinese & $>20$ & 8,000 & 2,000 & 1,563 \\
NISQA & English (German) & 5 (24) & 11,020 & 2,700 & 232 \\ \hline
\end{tabular}
\end{adjustbox}
\caption{Specifications of the three MOS prediction datasets}
\label{tab:datasets}
\end{table}

For data splitting, the BVCC dataset follows its predefined splits, while the Tencent corpus is split using a fixed random seed for consistency. On the NISQA dataset, we merge the live and simulated train splits for training and merge the live and simulated validation splits for validation. The model is tested on the German LiveTalk subset to introduce language diversity and evaluate the model's generalization performance. We independently train and evaluate the models on each dataset, focusing solely on utterance-level performance, as system-level scores are not available for all datasets.

\subsection{Method}

The Speech SSL models all operate on a 16~kHz sampling rate; therefore, all speech clips were downsampled to this corresponding frequency. To standardize input length, we either repetitively pad or randomly crop the audio samples to a duration of 8 seconds, depending on the original signal length. This ensures that all inputs fed into the attached projection head maintain a consistent shape.

\begin{figure}
    \centering
    \includegraphics[width=1\linewidth]{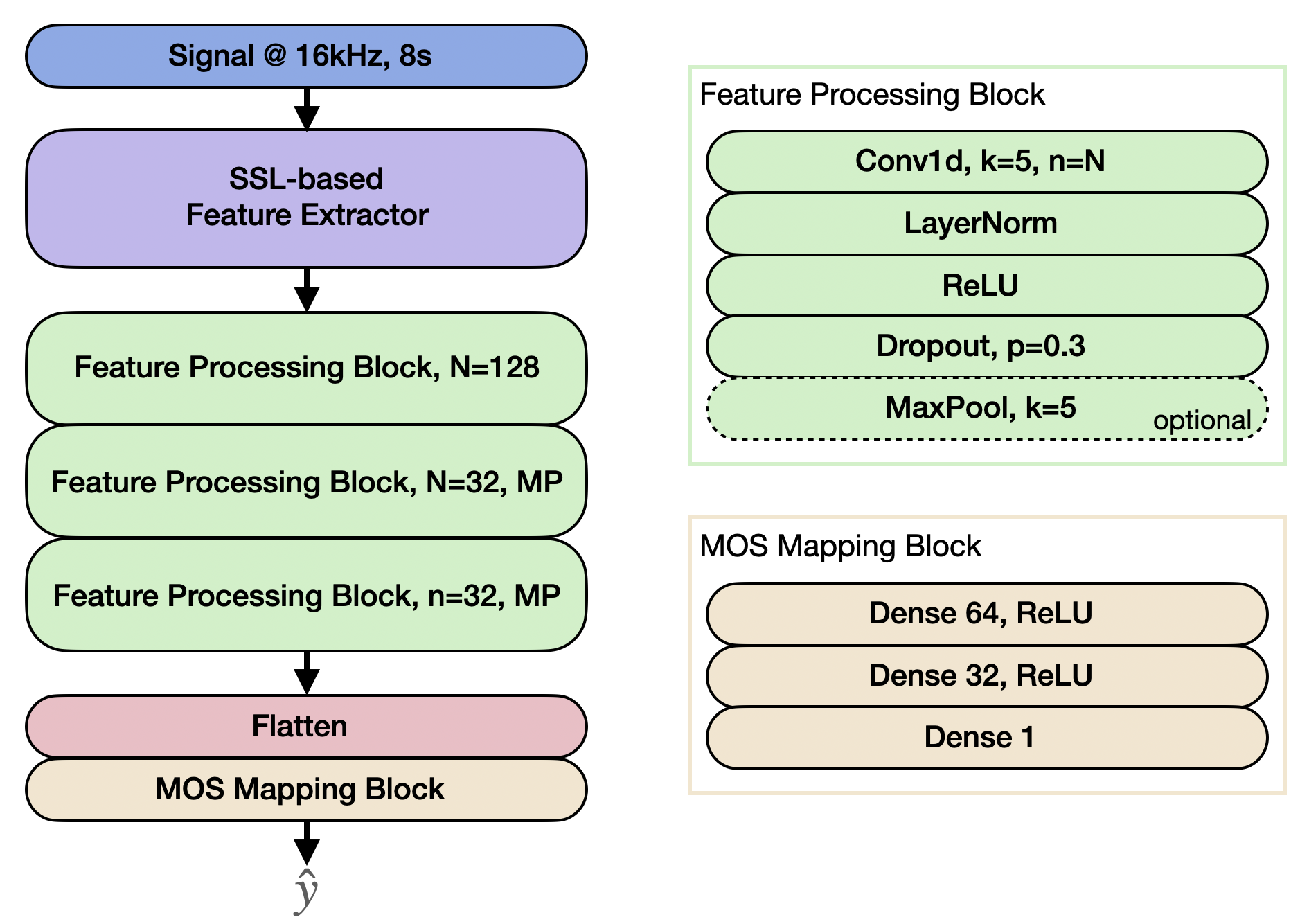}
    \caption{Architecture for the projection head}
    \label{fig:projection-head}
\end{figure}

We propose a novel design for the projection head model that maps the extracted features to MOS scores, motivated by the design of DNSMOS \cite{DNSMOS}. It features generalized compatibility with any feature extractor and can be applied to train on any MOS dataset. The detailed architecture of the projection head is depicted in Figure \ref{fig:projection-head}. For the features extracted from each layer of each model, the projection head first processes the input using 1-dimensional convolutional layers, followed by flattening, to map the features to a scalar MOS score. The use of 1-dimensional convolutional layers is intended to effectively handle the generally uncorrelated nature of SSL-extracted features along the channel axis, meanwhile minimizing parameter count.

Three commonly referenced performance metrics in this domain are mean square error (MSE), linear correlation coefficient (LCC) \cite{LCC}, and Spearman’s rank correlation coefficient (SRCC) \cite{SRCC}. Correlation-based measures, such as LCC and SRCC, are less sensitive to mismatches in value scales compared to MSE, making them more reliable. In our study, we prioritize LCC as the key performance metric, noting that it typically yields values similar to SRCC.

\subsection{Direct use of SSL models and their fine-tuning}
\label{sec:fine-tune}

We consider two situations in our experiments: (a) We directly use features from a selected layer of a chosen SSL model. (b) We fine-tune a chosen SSL model for MOS prediction, and then select a layer of the fine-tuned SSL model. The fine-tuning helps to achieve a better MOS prediction performance, while the position of the optimal layers remains unchanged. This will be shown experimentally later in section \ref{sec:experiments}.

For design simplicity, we fine-tune the model on MOS prediction as a classification task, by quantizing the MOS labels for each sample into values that are multiples of $0.5$. Since MOS values range from 1 to 5, this results in 9 classes. We fine-tune the model exclusively on the BVCC and Tencent datasets, as the NISQA dataset is used to test the generalization ability on out-of-domain data where training and test sets are in different languages. We evaluate the fine-tuned model on the validation split to avoid data leakage. We do not freeze the feature extractor, and the models are saved after each epoch.

\section{Experiments}
\label{sec:experiments}

We trained our projection head models for $30$ epochs using the Adam \cite{Adam} optimizer, with a learning rate of $10^{-4}$, a training batch size of $64$, and mean square error (MSE) loss as the training objective \footnote{Code and model checkpoints can be found in \url{https://github.com/Hope-Liang/SSL_Layer_MOS}.}. Model evaluation was conducted on the validation set after each epoch, and the final model was selected based on the highest LCC performance observed on the validation data. To account for variability in training, we report the average performance across five random runs for each scenario, i.e., each layer of each model. The training was performed on a single Nvidia A100 40~GB GPU, with each projection head model requiring 10-20 minutes to complete.

\begin{table*}[h!]
    \centering
    \begin{adjustbox}{width=\textwidth}
    \begin{tabular}{|c|c|c|ccc|c|ccc|c|ccc|}
        \hline
        \multirow{2}{*}{Model} & \multirow{2}{*}{\# Layers} & \multicolumn{4}{c|}{BVCC} & \multicolumn{4}{c|}{Tencent} & \multicolumn{4}{c|}{NISQA\_TEST\_LIVETALK} \\
        \cline{3-14}
        & & Best Layer & MSE & LCC & SRCC & Best Layer & MSE & LCC & SRCC & Best Layer & MSE & LCC & SRCC \\
        \hline
        \multicolumn{14}{|c|}{Attached with projection head, results reported as the average of five random trainings.} \\
        \hline
        w2v2\_base & 12 & 3 & 0.220 & 0.867 & 0.866 & 3 & 0.184 & 0.962 & 0.963 & 4 & 0.384 & 0.802 & 0.793 \\
        w2v2\_base\_ft\_best & 12 & 4 & 0.223 & 0.876 & 0.876 & 4 & 0.182 & 0.963 & 0.964 & - & - & - & - \\
        hubert\_base & 12 & 3 & 0.237 & 0.866 & 0.867 & 5 & 0.169 & 0.964 & 0.965 & 4 & 0.307 & 0.825 & 0.810 \\
        wavlm\_base & 12 & 3 & 0.229 & 0.864 & 0.864 & 4 & 0.173 & 0.963 & 0.963 & 4 & 0.269 & 0.863 & 0.855 \\
        w2v2\_xlsr\_300m & 24 & 5 & 0.225 & 0.884 & 0.884 & 7 & 0.166 & 0.972 & 0.972 & 5 & 0.207 & 0.920 & 0.909 \\
        w2v2\_xlsr\_1b & 48 & 7 & 0.223 & 0.885 & 0.884 & 15 & 0.166 & 0.974 & 0.974 & 43 & 0.226 & 0.912 & 0.900 \\
        w2v2\_xlsr\_2b & 48 & 7 & 0.210 & 0.886 & 0.886 & 13 & 0.157 & 0.974 & 0.974 & 39 & 0.198 & 0.922 & 0.917 \\
        \hline
        \multicolumn{14}{|c|}{Other state-of-the-art SSL-based models, results quoted from literature \cite{UTMOS}.} \\
        \hline
        UTMOS & - & - & 0.276 & 0.883 & 0.881 & - & - & - & - & - & - & - & - \\
        SSL-MOS \cite{SSL-MOS} & - & - & 0.380 & 0.869 & 0.871 & - & - & - & - & - & - & - & - \\
        \hline
    \end{tabular}
    \end{adjustbox}
    \caption{Performance metrics across different models and datasets.}
    \label{tab:model_results}
\end{table*}

\subsection{Direct use of SSL models}

\begin{figure}[h!]
    \centering
    \begin{subfigure}[b]{\columnwidth}
        \centering
        \includegraphics[width=0.6\columnwidth]{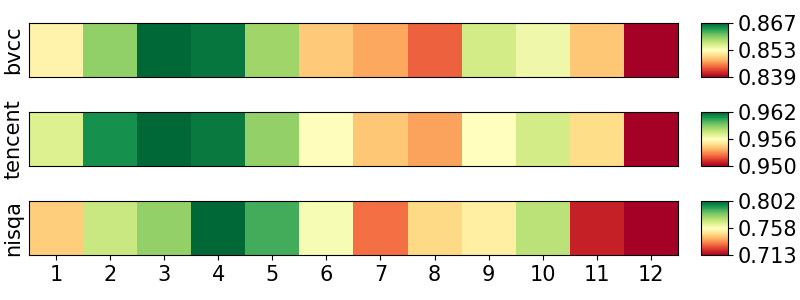}
        \caption{Wav2vec2\_base}
        \label{fig:grid_w2v2_base}
    \end{subfigure}

    \begin{subfigure}[b]{\columnwidth}
        \centering
        \includegraphics[width=0.6\columnwidth]{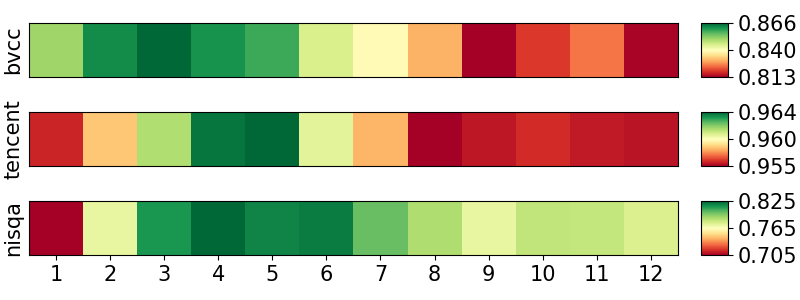}
        \caption{Hubert\_base}
        \label{fig:grid_hubert_base}
    \end{subfigure}

    \begin{subfigure}[b]{\columnwidth}
        \centering
        \includegraphics[width=0.6\columnwidth]{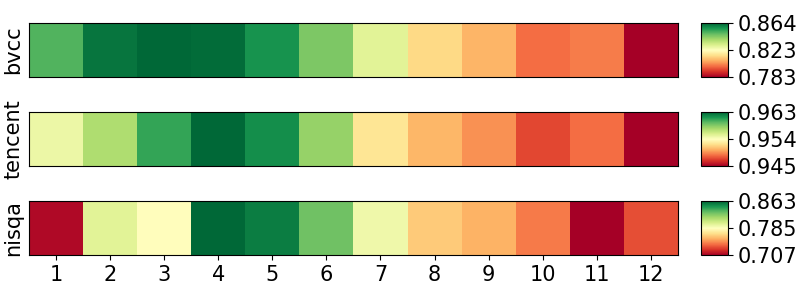}
        \caption{Wavlm\_base}
        \label{fig:grid_wavlm_base}
    \end{subfigure}
    \caption{LCC - Layer plots for base size models}
    \label{fig:arch_comp}
\end{figure}

We begin by comparing three base-size models from each speech SSL family, all trained on the 960-hour LibriSpeech dataset with similar parameter counts, each having 12 layers. Figure \ref{fig:arch_comp} shows a grid plot of LCC values for each model's layers across three test sets. The coloring schemes are with respect to the range of LCC values within each individual subplot, to highlight how performance varies across the layers.

A consistent trend emerges across datasets and models: the most suitable layers for MOS prediction are around layers 3 to 5, which are in the shallower to mid-level layers.

\begin{figure}[h!]
    \centering
    \begin{subfigure}[b]{\columnwidth}
        \centering
        \includegraphics[width=1.0\columnwidth]{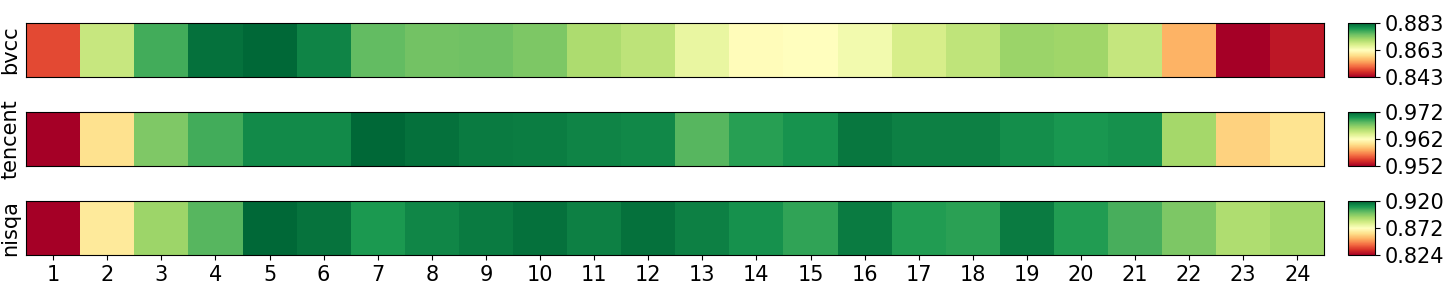}
        \caption{Wav2vec2\_xlsr\_300m}
        \label{fig:grid_w2v2_xlsr_300m}
    \end{subfigure}

    \begin{subfigure}[b]{\columnwidth}
        \centering
        \includegraphics[width=1.0\columnwidth]{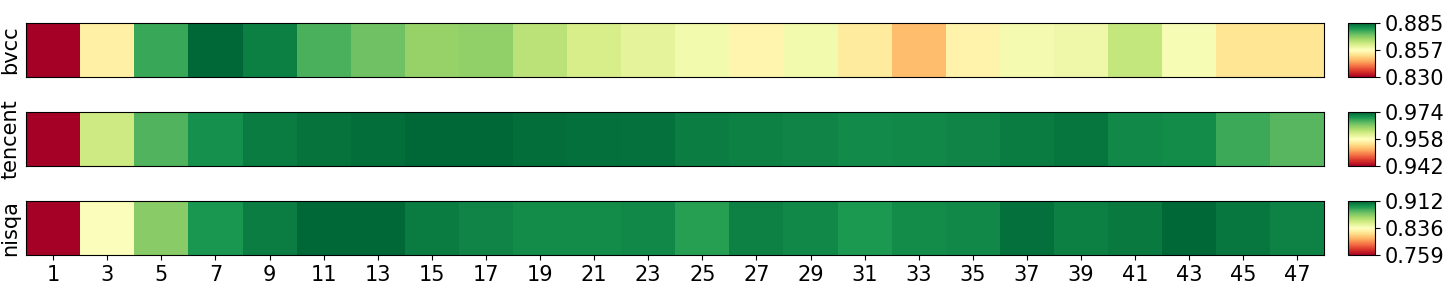}
        \caption{Wav2vec2\_xlsr\_1b}
        \label{fig:grid_w2v2_xlsr_1b}
    \end{subfigure}

    \begin{subfigure}[b]{\columnwidth}
        \centering
        \includegraphics[width=1.0\columnwidth]{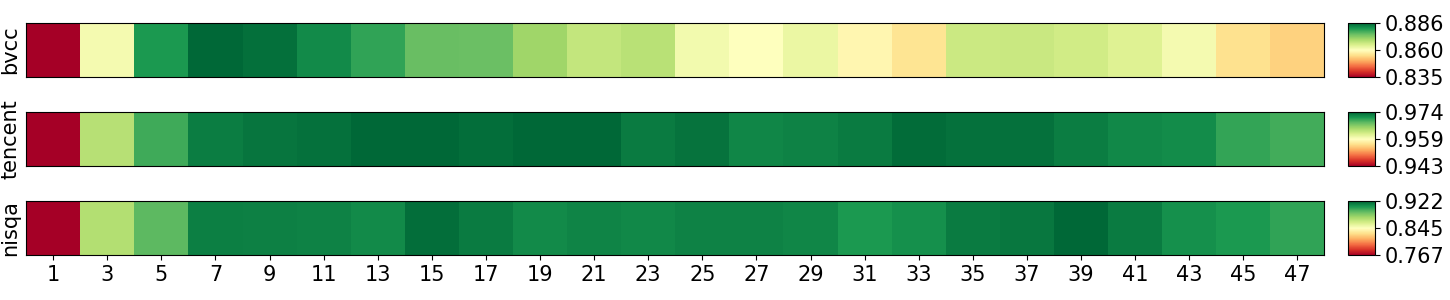}
        \caption{Wav2vec2\_xlsr\_2b}
        \label{fig:grid_w2v2_xlsr_2b}
    \end{subfigure}

    \caption{LCC - Layer plots for w2v2\_xlsr models}
    \label{fig:size_comp}
\end{figure}

Wav2Vec2 offers various pre-trained sizes available to address tasks of different complexities. It is generally assumed that larger models yield better performance on downstream tasks. To evaluate this, we test three larger variants of the model, pre-trained on 436,000 hours of multi-dataset data. We examine every layer of the 24-layer, 300M model and every other layer of the 48-layer, larger models.  We employ the same grid plot method as before, as shown in Figure~\ref{fig:size_comp}.

Across all models, the best-performing layers for the BVCC dataset appear in the early layers, while for the other datasets, performance remains high starting from layers just after the first few. We attribute this to the noisy labeling in the BVCC dataset, which increases the task's difficulty. In the other datasets, the large SSL model's extensive capacity appears to saturate the task of MOS prediction across multiple layers. This is further evidenced by the LCC values: the best layer for BVCC achieves around 0.88, while the best layers on the other datasets reach 0.974 and 0.922, with similarly high values across multiple layers. The language domain mismatch between training and testing splits we use in the NISQA dataset likely explains its relatively lower performance compared to Tencent.

\subsection{MOS prediction fine-tuned SSL models}

\begin{figure}[h!]
    \centering
    \begin{subfigure}[b]{0.48\columnwidth}
        \centering
        \includegraphics[width=\columnwidth]{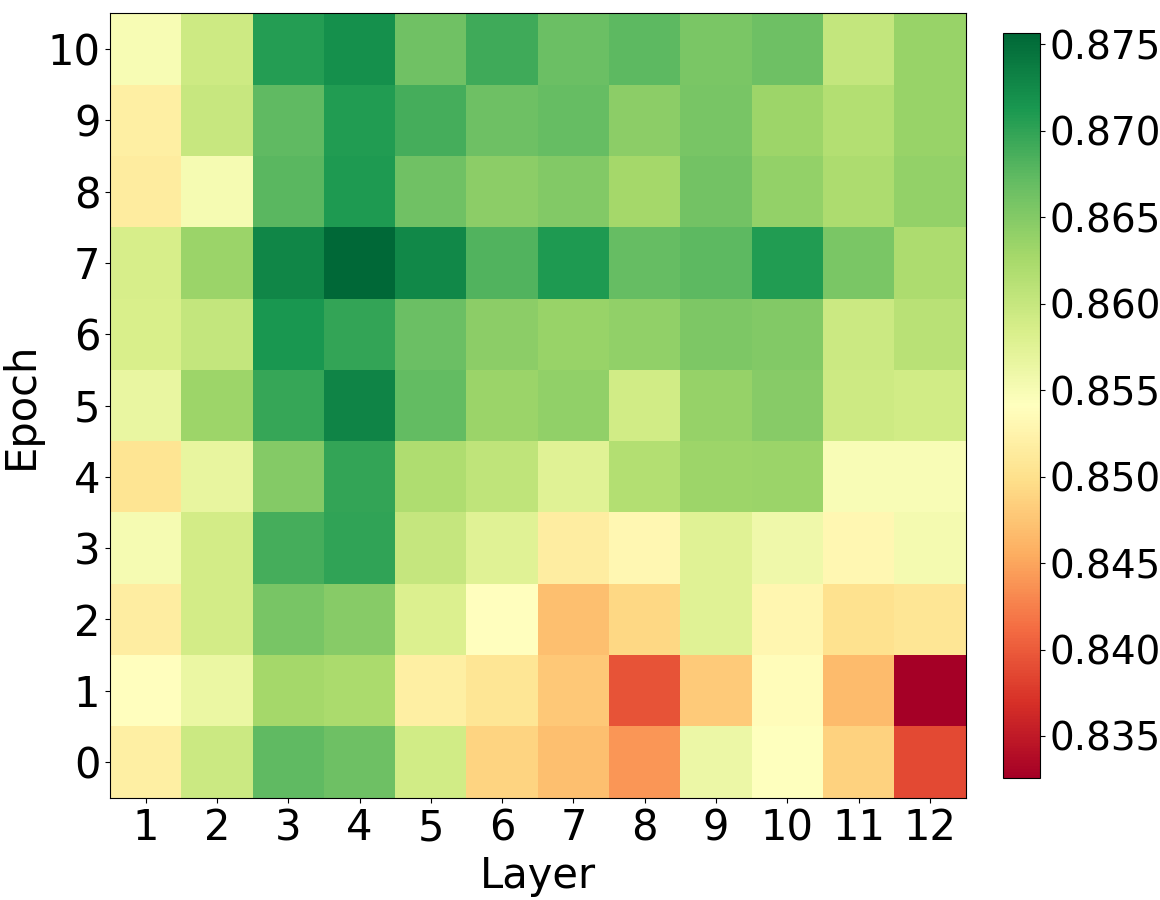}
        \caption{BVCC-finetuned}
        \label{fig:finetune-sub1}
    \end{subfigure}
    \hfill
    \begin{subfigure}[b]{0.48\columnwidth}
        \centering
        \includegraphics[width=\columnwidth]{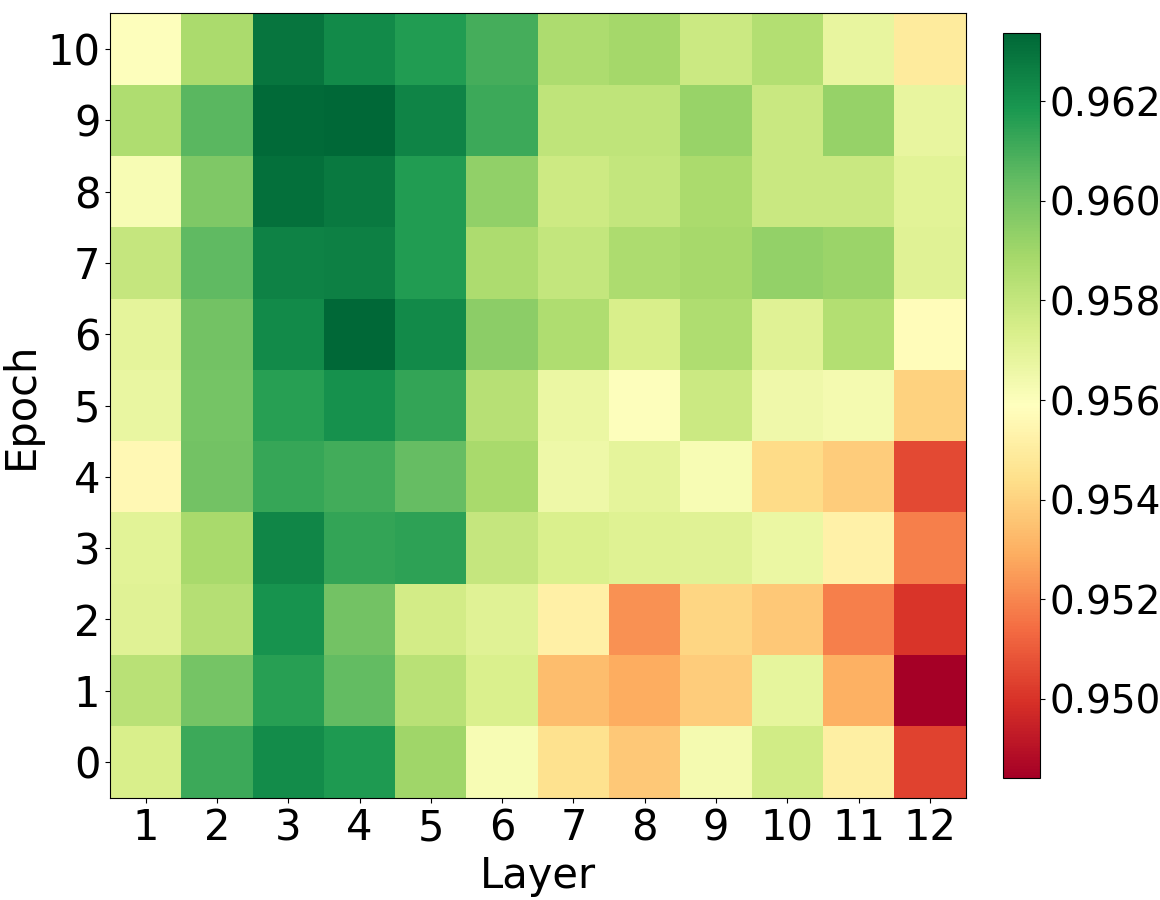}
        \caption{Tencent-finetuned}
        \label{fig:finetune-sub2}
    \end{subfigure}
    \caption{Wav2Vec2 fine-tuned and tested on MOS datasets}
    \label{fig:finetune_comp}
\end{figure}

We finetune the Wav2vec2\_base model as explained in section \ref{sec:fine-tune}. The model is fine-tuned on the training split of each dataset for 10 epochs, with a learning rate of $3e-5$, a batch size of $32$, and a warm-up ratio of $0.1$. We treat the model fine-tuned after each epoch as an individual model. We visualize the LCC values using a grid plot, shown in Figure~\ref{fig:finetune_comp}, where each row corresponds to a model training epoch and each column represents a layer. The LCC values are color-coded from red to green across the entire plot, with the colors indicating relative performance across all models and layers, rather than being independently scaled within each row.

We observe that fine-tuning generally enhances the latent features for the task of MOS prediction, as indicated by the progression of greener colors moving upwards in the plots. However, the most performant layer remains consistent with its position before fine-tuning, albeit with improved performance.

\subsection{Overall performance compared with SOTA methods}

To enable a vertical comparison of performance across different speech SSL feature extractors and benchmark them against SOTA methods, we present the results in Table \ref{tab:model_results}. For each model, we report the best-performing layer, identified by the highest LCC value, along with its performance across the three metrics mentioned.

It is evident (and not surprising) that, in general, larger feature extractor models yield better performance. The base-size models from Wav2Vec2, HuBERT, and WavLM exhibit similar performance on the BVCC and Tencent datasets. However, WavLM features demonstrate better generalization ability, as evidenced by its superior performance on the NISQA\_test\_livetalk dataset, where the training and testing data are in different languages.

The most effective layer for the MOS prediction task typically resides around the first quarter of the model’s total layers, regardless of the model's size or the amount of training data. Even after fine-tuning on the downstream MOS prediction task for a few epochs, the optimal layer's position remains consistent, albeit with slight performance improvements. For the NISQA\_test\_livetalk test set, the mismatch between the training and testing languages leads to performance saturation, with the layers around the first quarter showing only marginal differences compared to those reported in the table. Therefore, we believe the general trend holds true. Reminiscent of the layer-wise study in \cite{pasad2023comparative}, the most effective layer for the MOS prediction task is where the features exhibit high correlation with both spectrogram features and phonetic and semantic labels.

We also include two SOTA models, UTMOS and SSL-MOS, for comparison. Notably, a simple wav2vec2\_xlsr\_300m model with feature layer selection and a small projection head already surpasses UTMOS. UTMOS leverages multiple input features including speech SSL features from various models, phoneme sequences, and listener identity, yet it falls short in performance. For datasets without such extra information, the method becomes inapplicable.

\section{Conclusion}
\label{sec:conclusion}

Our study highlights the importance of layer selection in speech SSL models for predicting MOS. Contrary to the common practice of using features from the final layer, we find that early layers, typically around the first quarter of the SSL model, yield significant performance improvements. This trend holds across different SSL models, including Wav2Vec2, HuBERT, and WavLM, and remains consistent regardless of model size or training data. By employing a simple, parameter-efficient projection head attached to the optimal layers, we surpass SOTA models that use more complex designs and additional data. Our findings emphasize that top-layer dropping of speech SSL models can both improve system performance and efficiency.

\end{document}